\pdfoutput=1
\documentclass[a4paper,english]{article}
\usepackage[T1]{fontenc}
\usepackage[latin9]{inputenc}
\usepackage{color}
\usepackage{amstext}
\usepackage{amssymb}
\usepackage{graphicx}
\usepackage{setspace}
\usepackage{authblk}
\makeatletter

\providecommand{\tabularnewline}{\\}

\@ifundefined{date}{}{\date{}}
\usepackage{draftcopy}

\@ifundefined{showcaptionsetup}{}{%
 \PassOptionsToPackage{caption=false}{subfig}}
\usepackage{subfig}
\makeatother

\begin{document}

\title{ALPINE: A Bayesian System for Cloud\textcolor{blue}{{} }Performance Diagnosis and
Prediction}
\author[1]{Karan Mitra\thanks{Corresponding author. Website: http://karanmitra.me, Email:  karan.mitra@ltu.se}} 
\author[1]{Saguna Saguna}
\author[1] {Christer \AA{}hlund}
\author[2]{Rajiv Ranjan}

\affil[1]{Department of Computer Science, Electrical and Space Enginering, Lule\aa{} University of Technology, Skellefte\aa{}, Sweden}
\affil[2]{School of Computing Science, Newcastle University, United Kindgom}

\renewcommand\Authands{ and }


\maketitle
\begin{abstract}
Cloud performance diagnosis and prediction is a challenging problem
due to the stochastic nature of the cloud systems. Cloud performance
is affected by a large set of factors including (but not limited to)
virtual machine types, regions, workloads, wide area network delay
and bandwidth. Therefore, necessitating the determination of complex
relationships between these factors. The current research in this
area does not address the challenge of building models that capture
the uncertain and complex relationships between these factors. Further,
the challenge of cloud performance prediction under uncertainty has
not garnered sufficient attention. This paper proposes develops and
validates ALPINE, a Bayesian system for cloud performance diagnosis
and prediction. ALPINE incorporates Bayesian networks to model uncertain
and complex relationships between several factors mentioned above.
It handles missing, scarce and sparse data to diagnose and predict
stochastic cloud performance efficiently. We validate our proposed
system using extensive real data and trace-driven analysis and show
that it predicts cloud performance with high accuracy of 91.93\%. 
\end{abstract}


\section{Introduction}

\begin{singlespace}
Cloud computing through virtualization provides elastic, scalable,
secure, on-demand and cheaper access to computing, network, and storage
resources as-as-service \cite{ArmbrustEECS200928}. The cloud system
hides the complexity of managing these virtualized resources to provide
an easy way for the end users to deploy their applications on the
cloud. The rapid surge in demand for cloud computing in the recent
years has led to the emergence of several cloud providers such as
Amazon Elastic Compute Cloud (EC2) and Google Compute Engine (GCE).
CloudHarmony \cite{cloudharmony}, a major cloud provider comparison
website lists ninety-six such cloud providers. Most cloud providers
offer relatively similar functionality, albeit at different prices
and with different service level agreements. Although each cloud provider
aims to maximise their revenue by providing a broad range of applications
and services to the end users, the quality of service (QoS) offered
by them can differ substantially. The multi-tenant model inherent
in cloud systems, and the limitations posed by global Internet bandwidth
may cause differences in QoS provided by the cloud providers that
can hamper applications hosted on the clouds \cite{Leitner2016}. 
\end{singlespace}

Cloud performance benchmarking (regarding QoS), diagnosis and prediction
is a highly challenging problem \cite{Leitner2016,Ward2014}. Each
cloud provider may provide a complex combination of cloud service
configurations at various geographically distributed regions all over
the globe (in a cloud datacenter). These service configurations include
a plethora of virtual machine instance types, and network and storage
services. Zhang \emph{et al.} \cite{zhang15} note that Amazon Web
Service alone offers six hundred and seventy-four such combinations
differentiated by price, geographical region, and QoS. Each combination
of these services provided over the Internet may lead to QoS variations.
Therefore, it is imperative for the end users to monitor the QoS offered
by the cloud providers during and after selection of a particular
cloud provider for hosting their applications. 

Cloud performance monitoring and benchmarking is a widely studied
problem \cite{Ward2014,Alhamazani2015}. Recent research in this area
(e.g., \cite{li2013,AlhamazaniTCC15,workbench15}) has developed tools
and platforms to monitor cloud resources across all cloud layers,
i.e., Infrastrastrucure-as-a-Service (IaaS), Platform-as-a-Service
(PaaS), and Software-as-a-Service (SaaS). Further, recent research
(e.g., \cite{Leitner2016}) has also widely studied the performance
of several cloud platforms based on various applications, constraints,
and experimental setups \cite{Leitner2016}. However, the challenge
of performing root-cause diagnosis of cloud performance by critically
studying the effect of multiple influencing factors taken together
has not garnered sufficient attention. Further, the current research
does not deal with the challenge of handling uncertainty caused due
to the uncontrollable (hidden) factors prevalent in the stochastic
cloud environment. Lastly, the current research does not aim to build
a unifying model for cloud performance diagnosis and prediction. 

\emph{Our contribution: }This paper proposes, develops and validates
ALPINE, a systematic and a unifying system for cloud performance diagnosis
and prediction. ALPINE incorporates Bayesian networks to model uncertain
and complex relationships between several factors such as CPU type,
geographical regions, time-of-the-day, day-of-the-week, cloud type,
and the benchmark-type. Using Bayesian networks and the Expectation
Maximization algorithm, ALPINE handles missing, scarce and sparse
data to diagnose and predict stochastic cloud performance efficiently.
We validate ALPINE using extensive real data and trace-driven analysis
and show that it predicts cloud performance with high\textcolor{black}{{}
accuracy of 91.93\%}. 

The rest of the paper is organised as follows: Section 2 presents
the related work. Section 3 presents ALPINE. Section 4 presents
the results analysis. Finally, section 5 presents the conclusion and
future work.

\section{Related Work }

The problem of cloud performance monitoring, benchmarking and prediction
has got significant interest from both industry and academia \cite{amazoncloudwatch,cloudharmony,Ward2014,Varghese14,Alhamazani2015}.
There are already commercial and academic cloud monitoring and benchmarking
systems available in the cloud domain. For example, CloudHarmony \cite{cloudharmony}
provides cloud benchmarking, reporting and selection service based
on several parameters such as regions, latency and throughput. Amazon
EC2 provides CloudWatch \cite{amazoncloudwatch}, a cloud monitoring
service for monitoring virtual machine instances running on Amazon
EC2 clouds. CloudWorkbench \cite{workbench15} provides a Web-based
system for benchmarking IaaS clouds. However, these systems and methods
simply provide raw aggregated measurements and do not provide any
analysis and recommendations. 

The research work presented in this paper is motivated by \cite{Leitner2016}
where the authors present an in-depth analysis of the results regarding
performance variability in major cloud providers such as Amazon EC2
and Google AppEngine. Most importantly, the authors studied performance
variability and predictability of cloud resources by performing experimentation
for several days and by collecting real data traces. We used these
data traces in this paper. 

The work presented by \cite{Leitner2016} was limited based on several
factors. For instance, the authors did not critically determine the
influence of multiple factors taken together to ascertain the degree
of change that occurs when the values of these factors are varied.
Further, the authors did not develop a model that can be used to predict
cloud performance under uncertainty and missing data values. Compared
to the work presented in \cite{Leitner2016}, this paper presents
a systematic and unifying model based on Bayesian networks (BNs) to
model complex relationships between several factors for efficient
cloud performance diagnosis and prediction. 

Recently, BNs were applied in the area of cloud computing (e.g., \cite{Jaatun09raey,Bashar2013,Tang14}).
Bashar \cite{Bashar2013} use BNs for autoscaling of cloud datacenter
resources by balancing the desired QoS and service level agreement
targets. The author using preliminary studies show the BNs can be
utilised efficiently to model workloads, and QoS factors like CPU
usage and response time. However, they did not discuss in detail how
BNs can be created and validated by the stakeholders. Further, their
work was limited to simpler simulation studies and did not consider
realistic user workloads. Compared to the work presented by \cite{Bashar2013},
in this paper, we consider the challenge of efficient cloud performance
diagnosis and prediction considering major public Cloud providers
such as Amazon EC2 and Google AppEngine. 

Compared to the state-of-the-art research in the area \cite{Jaatun09raey,Bashar2013,Tang14,Leitner2016,Varghese14},
the main aim of this paper is to develop a system for critical diagnosis
and prediction of cloud performance under uncertainty. Our system,
ALPINE, considers several factors such as time-of-the-day, day-of-the-week,
virtual machine-type, regions and different types of benchmarks and
efficiently models complex relationships between these parameters
for cloud performance diagnosis and prediction. Using realistic data
provided by Leitner and Cito \cite{Leitner2016}, in this paper, we
show how the stakeholders can develop BNs to perform probabilistic
cloud performance diagnosis and prediction, and to determine the best
combination of cloud resources for a given QoS level.

\section{ALPINE: Bayesian Cloud Performance Diagnosis and Prediction}

This section presents ALPINE - a Bayesian system for cloud QoS diagnosis
and prediction. Fig. 1. shows our high-level approach. As can be observed
from this figure, first, benchmark data is collected by the stakeholders
through experimentation or via third-party services such as Cloud
Workbench \cite{workbench15} and CloudHarmony \cite{cloudharmony}.
Second, this data is pre-processed and is stored in a database. Third,
a Bayesian Network (BN) is learned using the pre-processed data or
is manually created by the domain expert. In the case of manual BN
creation, the model is created using domain expert's knowledge/experience;
or it is learned using the pre-processed data which is then carefully
calibrated by the domain expert. Fourth, the modelled
BN is then used for probabilistic diagnosis by entering the evidence
in the form of probability assignment, i.e., a likelihood of a random
variable (or facrtor) taking a particular value is determined by introducing
evidence into the BN (discussed later in detail). Fifth, if the diagnostic
results are deemed to be sufficient, this BN can be used by the stakeholders
for both diagnosis and prediction, and for actual usage; else, steps
one to three are repeated to develop the best BN.

\subsection{Modelling Bayesian Networks for Cloud QoS Diagnosis and Prediction}

We consider Bayesian Networks (BNs) for cloud QoS diagnosis and prediction.
We selected BNs over Fuzzy Logic, Neural Networks and Decision Trees
as a method based on its several advantages. These include: BNs learn
efficiently from scarce and sparse data. BNs deal effectively with
uncertainty in stochastic environments (such as clouds and networks).
BNs handle both numerical and categorical data. BNs can incorporate
domain knowledge. BNs do not require explicit rules to reason about
factors. BNs can be extended to dynamic Bayesian networks to reason
about several hypotheses over time. Finally, they can be used with
utility theory to make decisions under uncertainty \cite{russelandnorvig,Caqoem15}.
We now show how BNs can be used to model several parameters for efficient
for cloud performance diagnosis and prediction. A BN can be defined
as follows: 
\begin{figure}
\centering{}\includegraphics[width=1\columnwidth]{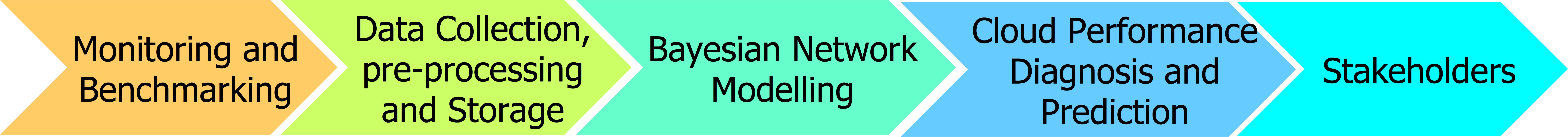}\caption{A Bayesian Network for cloud QoS diagnosis and prediction. }
\end{figure}

\textbf{Definition 1.} \emph{A Bayesian network (BN) is a directed
acyclic graph (DAG) where, random variables form the nodes of a network.
The directed links between nodes form the causal relationships. The
direction of a link from X to Y means that X is the parent of Y. Any
entry in the Bayesian network can be calculated using the joint probability
distribution (JPD) denoted as: 
\begin{equation}
P(x_{1},...,x_{m})=\prod_{i=1}^{m}P(x_{i}|Parents(X_{i}))\;\blacksquare
\end{equation}
where, $parents(X_{i})$, denotes the specific values of $Parents(X_{i})$.
Each entry in the joint distribution is represented by the product
of the elements of the conditional probability tables (CPTs) in a
BN }\cite{russelandnorvig}\emph{.} 

BNs provide a natural and a complete description of the problem domain;
it provides a rich description of the causal relationships between
several nodes (representing factors) in the BN model \cite{russelandnorvig}.
Fig. 2 shows example BNs for cloud QoS diagnosis and prediction. In
these BNs, the oval nodes represent the random variables that are
modelled together to determine their effect on each other probabilistically.
In a BN, the direction of an arc from one node(s) to another node(s)
denotes a parent-child relationship, where the parent node directly
affects the child node probabilistically. For example in Fig. 2 (d),
the arcs from the nodes \textquotedblleft Regions\textquotedblright{}
and \textquotedblleft Virtual Machine Size\textquotedblright{} towards
\textquotedblleft CPU\textquotedblright{} denote that these nodes
are the parents of the child node \textquotedblleft CPU\textquotedblright ;
and will be used to determine the effect of regions and virtual machine
size on the types of CPU used.
\begin{figure}
\begin{centering}
\subfloat[Naive Bayes' Network (NBN).]{\centering{}\includegraphics[scale=0.03]{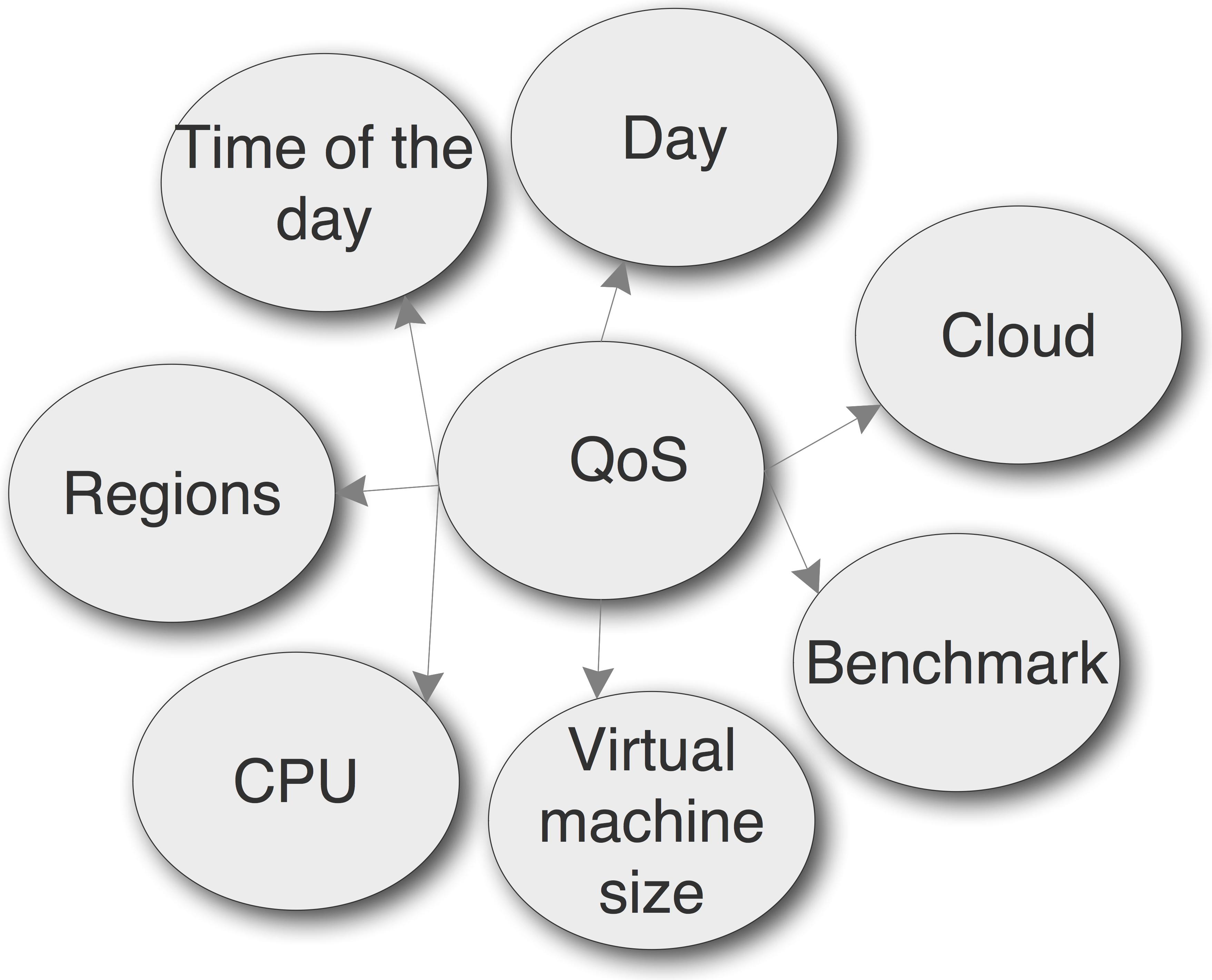}}\subfloat[Tree-Augmented Naive Bayes' Network (TAN).]{\begin{centering}
\includegraphics[scale=0.03]{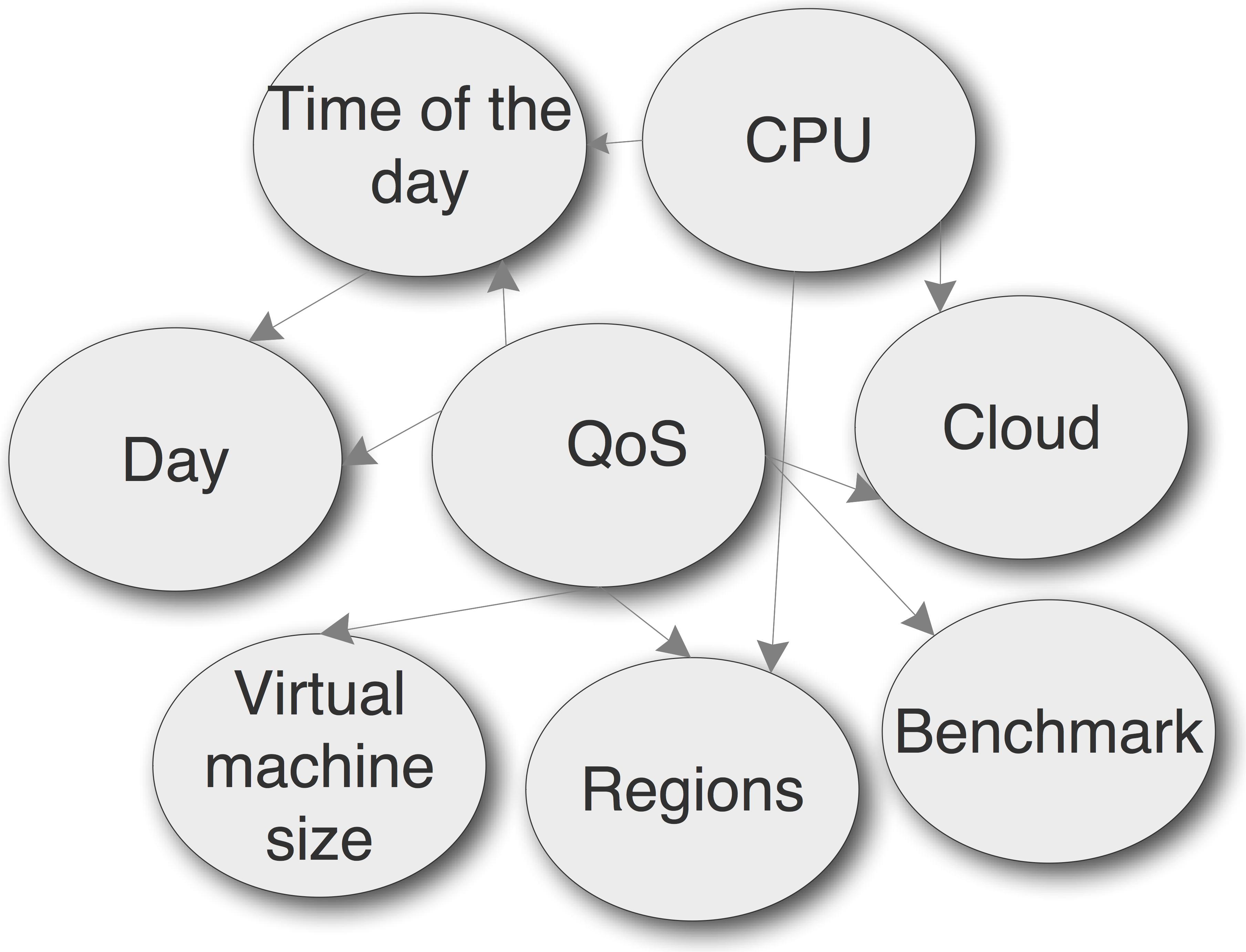}
\par\end{centering}

}
\par\end{centering}

\centering{}\subfloat[Noisy-Or Network (NOR).]{\begin{centering}
\includegraphics[scale=0.03]{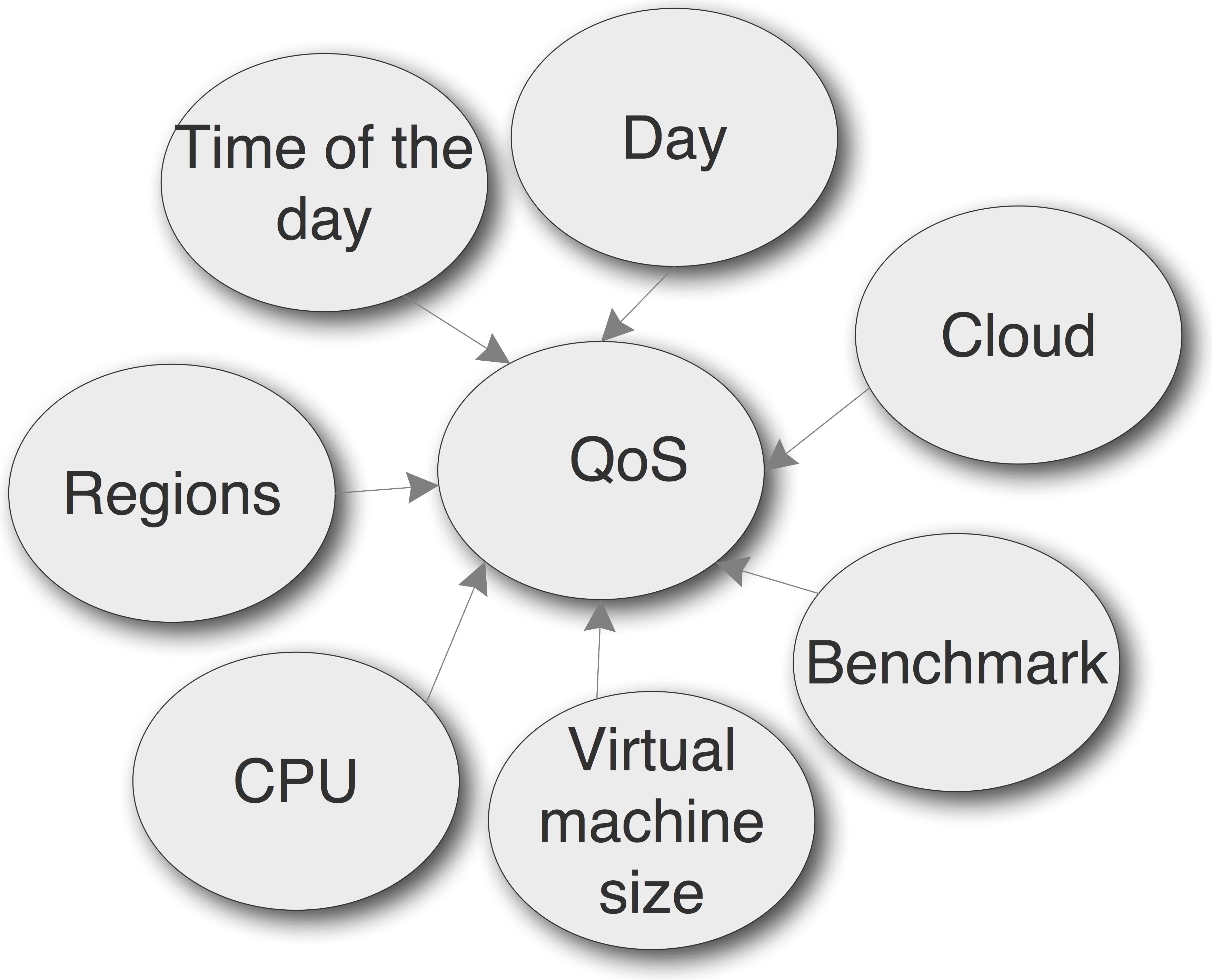}
\par\end{centering}

}\subfloat[Complex Bayesian Network (CBN).]{\begin{centering}
\includegraphics[scale=0.03]{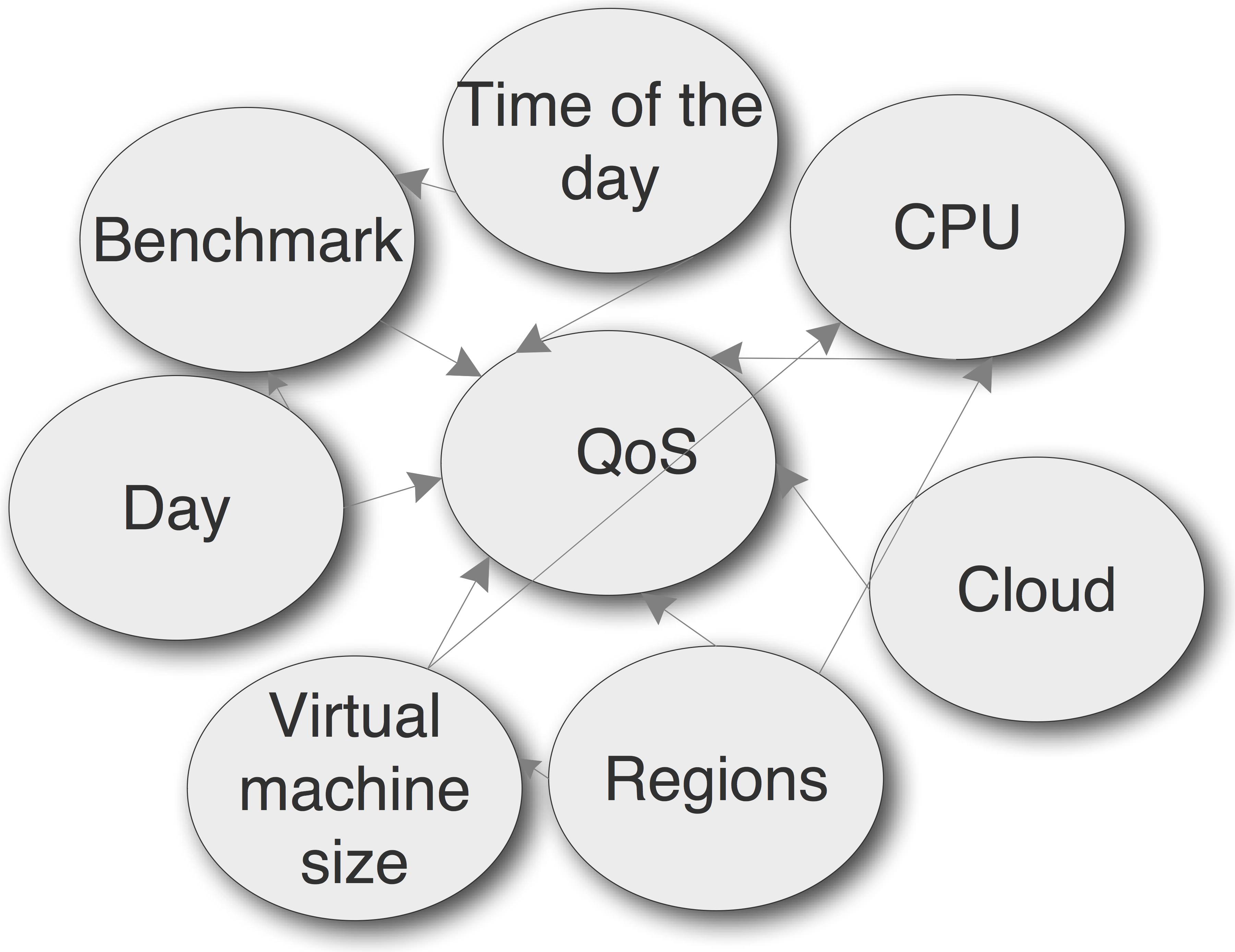}
\par\end{centering}

}\caption{Bayesian Networks for cloud QoS diagnosis and prediction. }
\end{figure}

A BN can be created in a number of ways\textcolor{black}{{} (see Fig.
2 (a) to Fig. 2 (d))} such as a Naive Bayes' Network, Noisy-Or Network,
Tree-Augmented Naive Bayes' Network, or a more complex model (such
as in Fig. 2 (d) where most of the nodes are connected to each other)
based on the principle of causality. Manual BN creation can be challenging
as the causal relationships can be hard to define by the stakeholders/domain
experts. To test the causal relationship between two factors or random
variables, consider the nodes \emph{A} and \emph{B}. Assume that the
domain expert fixes (assign probabilities) one of the state of node
\emph{A}\textcolor{black}{{} ($s\,\in\,S$ where $S$ is a set of states)},
to infer the states of node \emph{B}. Upon inference, if the states
of node \emph{B} do not change (degree of belief or probability of
a state $s\,\in\,S$ where $S$ is a set of states belonging to \emph{B}),
then the node \emph{A} is not a cause of node \emph{B}; otherwise
it is. For the sake of brevity, in the paper, we do not discuss various
methods for manual BN creation. The interested readers may refer to
\cite{russelandnorvig}. 

\textcolor{black}{Each node in a BN represents a random variable (RV
or factor in our case). This RV can be discretized into a number of
states $s\,\in\,S$. The $S$ is then assigned probabilities that
are represented via the conditional probability Table (CPT). In the
case of a continuous RVs, conditional probability distribution (CPD)
is defined that can take any distribution; for example, Gaussian distribution.
The CPT for each RV can be learned using a dataset or can be set by
the domain expert. As mentioned previously, setting the CPTs can be
quite challenging even if robust statistical methods are used \cite{Caqoem15}.
In such cases, the methods that consider maximum entropy can be used.
To create a BN automatically, stakeholders can also consider BN structural
learning algorithms such as structural expectation maximization and
Markov Chain Monte Carlo \cite{russelandnorvig}. For simplicity,
let's assume a BN shown in Fig. 2 (d). In this paper, we show that
even simpler BNs can be used efficiently to model, diagnose and predict
cloud QoS.}

Cloud QoS is stochastic and can be influenced by $N$ number of factors.
Further, each $n\,\in\,N$ can have $m\,\in\,M$ number of states.
In a BN, all the states can be inferred together by entering the evidence
$e\,\in\,E$ in the network which is not possible in other methods
such as regression analysis, decision trees, and neural networks.
By entering the evidence in a BN, we mean assigning a degree of belief
(associating probability) to a particular state $s\,\in\,S$ belonging
to an RV. For example, consider a BN as shown in Fig. 2 (d). To determine
the cloud QoS or ``QoS Value'' using the RV ``Cloud'', the stakeholder
can enter evidence into ``Cloud'' RV such as $P(``Cloud=aws''=1)\land P(``Cloud=gce''=0)$
to depict the degree of belief that for a particular ``QoS Value'',``aws''
``Cloud'' should be considered. Similarly, the probability of occurrence
of each $s\;\in S$ for all RVs can be entered as evidences $e\,\in\,E$
to determine the probability $\forall\,S$ for ``QoS Value'' RV. 

Once a BN is created via structural learning algorithms or by the
domain experts, they need to be validated. Usually, cross-validation
is performed to check the correctness and accuracy of the BN \cite{russelandnorvig}.
In cross-validation, a part of the training data is is to train/learn
the BN. The rest of the data or the test data is used to check model's
prediction accuracy. For BN model parameter learning, we consider
the most widely used Expectation-Maximization algorithm \cite{russelandnorvig}.
Once the stakeholders or domain experts are satisfied by BNs prediction
accuracy, these BNs can be utilised in the real-world use cases.

\section{Results Analysis}

This section presents the results related to ALPINE. We validate ALPINE
using GeNie Bayesian Network development environment \cite{genie}
as well as a realistic cloud benchmark dataset recently collected
by Leitner and Cito \cite{Leitner2016}. We chose this dataset based
on the fact that it is recent, comprehensive and covers a broad range
of factors that may affect the performance of clouds regarding communication,
I/O and storage.

\subsection*{Dataset}

The cloud benchmark dataset \cite{Leitner2016} contains 30,140 unique
records based on the data collected for one month regarding Amazon
EC2 (AWS) and Google Compute Engine (GCE) in the United States and
Europe regions. In particular, this dataset contains records related
to five benchmarks, namely, CPU, MEM, Compile, I/O, and OLTP. The
CPU benchmark was used to benchmark the compute capacity of the instance
(running in Amazon of Google data centers) by computing the total
time taken (in seconds (secs)) to check 20,000 natural numbers for
primeness. The MEM benchmark was used to measure the read-write memory
speed in MB/s by allocating 64 MB arrays in memory and copy one array
to the other fifty times. The Compile benchmark was used to measure
total cloning (from Github) and compilation time (in seconds) of the
jCloudScale Java program using the OpenJDK 7.0 toolkit. The I/O benchmark
was used measure (in Mb/s) the average disk read/write speed, computed
by reading and writing a 5 GB file for three minutes. Finally, OLTP
benchmark was used to measure the average number of queries per second
(queries/sec).

Table 1 shows the statistics related to all QoS values. We note that
this dataset does not contain MEM QoS values for GCE. Further, nearly
all QoS values are widely distributed. We now show that even with
variability in this dataset, ALPINE can efficiently diagnose and predict
cloud QoS. 

For cloud QoS diagnosis, we considered several BNs, such as a simple
Naive Bayes Network (NBN), Tree-augmented Naive Bayes Network (TAN),
Noisy-Or network (NOR), and a complex BN (CBN) as shown in Fig. 2.
We created the first two BNs automatically from the dataset. The latter
two BNs were created using expert's knowledge (by the authors). These
BNs comprise six random variables or BN nodes depicting eight different
factors present in the dataset. These include CPU, VM size, regions,
cloud providers, type of benchmark, time-of-the-day, day-of-the-week,
and QoS values. Except QoS value factor, all other factors were categorical,
ranging from two to eleven states $(s\in S)$. 
\begin{table}
\caption{Statistics related to all QoS values present in the dataset $(\text{\ensuremath{\Theta}})$.}

\centering{}%
\begin{tabular}{|c|c|c|c|c|c|}
\hline 
QoS Para. & Min. & Max. & Mean & Std. Dev. & Count\tabularnewline
\hline 
\hline 
CPU & 8.41 & 132.08 & 46.89 & 38.90 & 6894\tabularnewline
\hline 
Compile & 0 & 2654.5 & 230.07 & 171.50 & 7319\tabularnewline
\hline 
Memory & 611.65 & 6316.1 & 4114.5 & 1692.7 & 4581\tabularnewline
\hline 
I/O & 1 & 1009.6 & 17.96 & 51.11 & 7377\tabularnewline
\hline 
OLTP & 15.38 & 1130.25 & 310.05 & 281.74 & 3969\tabularnewline
\hline 
Combined & 0 & 6316.1 & 737.19 & 1584.2 & 30140\tabularnewline
\hline 
\end{tabular}
\end{table}

\subsection{CPU performance diagnosis}

The CPU benchmark aims to study the performance of hardware-dominated
applications hosted on the clouds. In particular, it seeks to examine
the effect of instance processing speed of cloud providers on the
hosted applications (task completion time in seconds). For this, we
studied several hypotheses using ALPINE. For instance, using a BN,
we studied the impact of several factors including the instance type,
time-of-the-day, day-of-the-week, region and CPU type on the applications'
task completion time. \emph{Using the same BN, we can not only determine
the impact these factors on the QoS value, but also each other. }For
example, we can easily answer the question that \emph{\textquotedblleft for
a certain QoS value, what is the most likely instance type, CPU type
and the region?\textquotedblright{}} i.e., using a single factor (CPU\_type),
we can infer the states of other factors (VM\_size, CPU\_type and
the region). Using a BN, we can infer the \emph{hidden truth} (phenomena
that cannot easily be explained by statistical tests) that may be
masked by traditional statistical tests. Most importantly, using probabilistic
analysis, experts can also use their intuition (i.e., they can assign
probabilities to particular states in a BN. For example, a state region
can be \textquotedblleft us\textquotedblright{} and \textquotedblleft eu\textquotedblright )
to reach several conclusions by studying several hypotheses. Traditional
statistical methods and the methods presented in \cite{li2013,zhang15,Varghese14,Leitner2016}
lack this capability.

The CPU dataset $(\theta_{(cpu)})$ contains 6894 data points for
both ``aws'' and ``gce'' clouds. We discretised the QoS values
into a ten states using hierarchical discretisation and by manual
fine tuning as shown in Table II. To study the impact of several factors
on the \emph{QoS value}, we first selected \textquotedblleft us\textquotedblright{}
\emph{region}, \textquotedblleft aws\textquotedblright{} as the cloud
provider (\emph{cloud}), and varied the \emph{VM\_size} as \textquotedblleft micro\textquotedblright ,
``small\textquotedblright , ``large\textquotedblright . These selections
were entered as evidence ($e\in E$) in a BN. For probabilistic inference,
this can be written as: P (\emph{QoS value}) = P (\emph{QoS Value}
| \emph{region} = ``us\textquotedblright , \textquotedblleft \emph{cloud}\textquotedblright{}
= \textquotedblleft aws\textquotedblright , \emph{VM\_size}= ``micro\textquotedblright ).
Through Bayesian analysis, we found clear differences offered by different
VM sizes. For instance, we found that for \emph{VM\_size}= ``small'',
there is 87\% chance (probability) that the task will be completed
between 82 and 103 seconds (state 9). Further, there is 86\% chance
that \emph{cpu} = \textquotedblleft Intel Xeon 2650v0 2.0 GHz'' will
be used. As expected, the ``large'' \emph{VM\_size} provided the
best performance. 

We concluded that for the ``large'' \emph{VM\_size}, there is 100\%
chance that the task will be completed between 11 and 20 seconds (state
2), offering up to five times better performance than ``small''
\emph{VM\_size}. Further, we note that ``aws'' \emph{cloud} only
uses the Intel Xeon 2760v2 2.50GHz CPU for providing predictable performance.
To our surprise, we found out that in the case of ``aws'' the \emph{``micro''}
\emph{VM\_size} provided significantly better CPU performance than
the ``small'' \emph{vm\_size}. In that, there is more than 84\%
chance that the task will be completed between 39 to 54 seconds (state
5), leading us to believe that a ``micro'' \emph{vm\_size} offers
two times better compute performance than the ``small'' \emph{vm\_size}.
Fig. 3 shows the screenshot of this case implemented in the GeNIe
platform. It is worth noting that for both ``small'' and ``micro''
\emph{vm\_size} mostly (84.5\% chance) use an ``Intel Xeon 2650v0
2 GHz'' CPU in the case of ``aws'' \emph{cloud} in the ``us''
\emph{region}. We then tested this hypothesis for the EU datacenter
and found similar results.\textcolor{black}{{} The $\theta_{cpu}$ also
contains values for ``ioopt'' and ``cpuopt'' specialised instances
for providing CPU and I/O optimised performance for ``aws'' }\textcolor{black}{\emph{cloud}}\textcolor{black}{,
respectively. After BN diagnosis, we found out that the ``ioopt''
}\textcolor{black}{\emph{VM\_size }}\textcolor{black}{provides the
best performance regarding}\textcolor{black}{\emph{ QoS\_value }}\textcolor{black}{and
with higher degree of certainty. In this case, all the }\textcolor{black}{\emph{QoS\_value
}}\textcolor{black}{lie below 11 seconds. On the other hand, and to
our surprise, the ``cpuio'' }\textcolor{black}{\emph{VM\_size }}\textcolor{black}{provides
nearly the same performance as the ``large''}\textcolor{black}{\emph{
VM\_size.}}

Finally, we studied the impact of several parameters on the \emph{QoS
value} for ``gce''. We found that ``gce'' provides highly predictable
results compared to ``aws'', and offers easily distinguishable performance
with different \emph{VM\_size}. Considering the ``micro'' \emph{VM\_size},
we found that there was greater than 94\% chance that the task completion
time was more than 103 seconds for both ``eu'' and ``us'' \emph{region}.
This result shows that ``aws'' ``micro'' \emph{VM\_size} provides
significantly better performance than ``gce'' ``micro'' \emph{VM\_size}.
On the other hand, we found that GCE\textquoteright s ``small''
\emph{VM\_size} performs at least three times better than ``aws''
``small'' \emph{VM\_size} with 100\% chance that the task completion
time would be between 20 to 32 seconds, compared to ``aws'' task
completion time of 82 to 103 seconds with approx. 87\% chance. In
THE case of the ``large'' \emph{VM\_size}, ``gce'' and ``aws''
performs similarly, offering task completion times between 11 to 20
seconds. It\textquoteright s also worth noting that ``gce'' always
selects the same processors for similar \emph{VM\_size} in ``eu''
and ``us'' \emph{region} leading to extremely high predictable CPU
performance compared to AWS. For example, ``gce'' always selects
the ``Intel Xeon 2.60 GHz'' processor for predicable performance
in both ``us'' and ``eu'' data centers or large VMs. We also studied
the impact of \emph{time} and \emph{day\_of\_the\_week} on \emph{QoS\_Value}
and found that these parameters do not significantly affect the CPU
performances.
\begin{table}
\caption{QoS value states representation using hierarchal discretization for
$\theta_{cpu}$. }

\begin{centering}
\begin{tabular}{|c|c|c|}
\hline 
State & Range & Counts\tabularnewline
\hline 
\hline 
1 & 0 to 11 & 480\tabularnewline
\hline 
2 & 11 to 20 & 2400\tabularnewline
\hline 
3 & 20 to 32 & 1092\tabularnewline
\hline 
4 & 32 to 39 & 31\tabularnewline
\hline 
5 & 39 to 54 & 916\tabularnewline
\hline 
6 & 54 to 61 & 3\tabularnewline
\hline 
7 & 61 to 67 & 50\tabularnewline
\hline 
8 & 67 to 82 & 87\tabularnewline
\hline 
9 & 82 to 103 & 885\tabularnewline
\hline 
10 & \emph{greater than} 103 & 950\tabularnewline
\hline 
\end{tabular}
\par\end{centering}

\end{table}
\begin{figure}
\centering{}\includegraphics[scale=0.15]{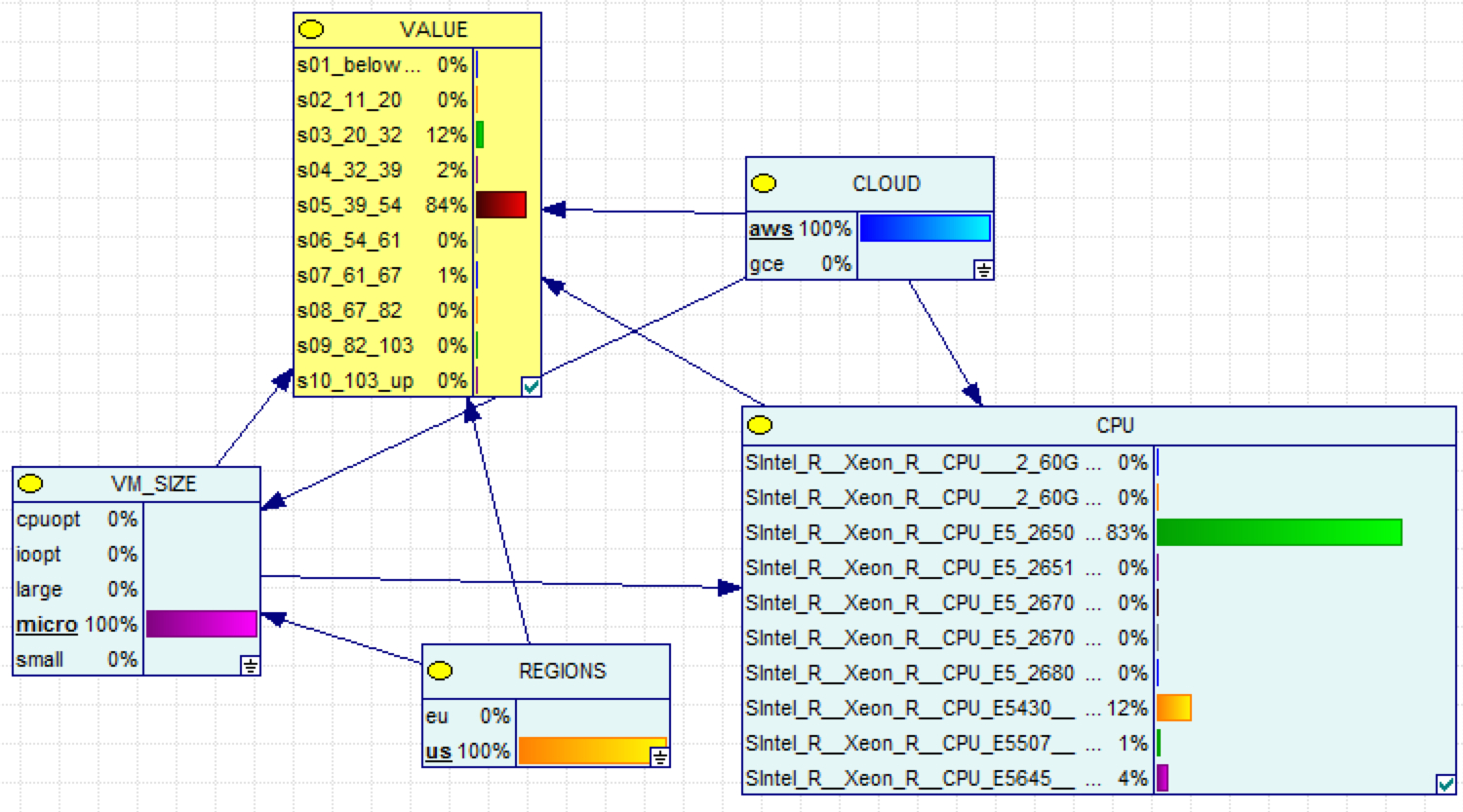}\caption{Screenshot of ALPINE implemented in GeNIe platform.}
\end{figure}

\subsection{Compile Diagnosis}

The aim of the compile benchmark is to study application's performance
on the clouds. Therefore, using Bayesian diagnosis, we studied the
impact of several factors mentioned above on the applications\textquoteright{}
compile time. As can be observed from Table I, the Compile dataset
$(\theta_{compile})$ contains a total of 7319 data points, representing
the QoS values for both ``aws'' and ``gce'' clouds. We discretised
the QoS values into a fifteen states using hierarchical discretisation
and by manual fine tuning as shown in Table III. We first analyzed
the performance of the ``aws'' cloud by varying the aforementioned
parameters. For example, by selecting the ``micro'' ``VM\_size''
in both ``eu'' and ``us'' regions, we found the QoS values to
the less predictable in the ``us'' region. In that, we found that
there is approx. 70\% chance that the QoS values will lie between
41 and 233 secs.; around 6\% chance that these values will lie between
233 and 405 secs; and 19\% chance that these values will lie between
405 and 701 secs. However, the ``micro'' ``vm\_size'' provides
more predictable performance in the ``eu'' ``region'' where there
is approx. 85\% chance that the QoS values will lie in the range of
4 and 233 seconds, and there is 8\% and 6\% chance that these values
will lie in the range of 233 to 405 seconds, and between 405 and 701
seconds, respectively.

The variation in the performance predicability can be attributed to
the fact that in both ``regions'', ``aws'' employs several different
``cpu types'' with varying probability. However, in the ``eu''
region, ``aws'' selects one of the CPU (``Intel 2650 2 Ghz'' processor)
in majority of the cases (with 84\% probability) compared to the ``us''
``region'' where there is 72\% chance that the same ``cpu'' will
be used. We also studied the performance of other VM types. When we
selected the ``small'' ``VM\_size'', the performance decreased
slightly but it becomes highly predictable (compared to ``micro''
VM\_size) with a 92\% chance that the QoS values will lie between
233 and 405 secs. We observed the similar behavior for both the regions. 

We then selected the ``large'' ``vm\_size'' and found that it
performed better than both ``micro'' and ``small'' instances.
In particular, we found that there was 97\% chance that the values
will lie between 41 and 233 secs for both the regions. For a thorough
diagnosis, we also studied the impact of optimised ``vm\_size''
such as, ``ioopt'' and ``cpuopt'' on the applications' performance.
As mentioned previously, these instances are optimised for I/O and
CPU operations and should offer better and more predicable performance
than the ``micro'', ``small'' and ``large'' ``vm\_size''.
For instance, we diagnosed that the ``ioopt'' ``VM\_size'' offers
better QoS values (with compile time lower than 112 seconds) with
92\% probability. Further, the ``cpuopt'' \emph{VM\_size} also provides
high QoS values with compile times in the range of 41 and 233 secs
with 97\% probability for the ``eu'' \emph{region}. There were no
QoS values present in the $\theta_{compile}$ dataset for the ``us''
region. We also found similar performance for the ``cpuopt'' instance
as well. \textcolor{black}{From our diagnosis we found it interesting
to note that the performance of the ``cpuopt'' and ``ioopt'' ``vm\_size''
is similar to the ``large'' }\textcolor{black}{\emph{VM\_size}}\textcolor{black}{.
This leads us to believe that instead of paying for ``cpuopt'' and
``ioopt'' VM\_size, ``large'' instance can be selected at lower
costs.}\textcolor{blue}{{} }We also studied the performance of all the
\emph{VM\_size} by also varying factors ``day-of-the-week'' and
``time-of-the-day'' and found no evidence that these factors''
significantly affect the QoS values for this benchmark for ``aws''
cloud.

Finally, we also diagnosed the performance offered by the ``gce''
cloud present in both ``us'' and ``eu'' \emph{regions}. In the
case of ``micro'' \emph{VM\_size}, there is approx. 91\% chance
that the QoS values will lie in the range of 405 and 701 secs. (state
4) in both ``eu'' and ``us'' \emph{regions}. Further, there is
approx. 99\% chance that the QoS values will lie in the range of 41
to 233 secs (state 2) for ``small'' \emph{VM\_size} in both ``us''
and ``eu'' \emph{regions}. It is also interesting to note that ``gce''
always selects the same\emph{ CPU} for similar \emph{VM\_size} compared
to ``aws'' cloud where different \emph{CPU} types can be selected
by the ``aws'' for same \emph{VM\_size}. In this dataset, there
were no data points for ``cpuopt'' and ``ioopt'' \emph{VM\_size}
therefore, we could not study the optimised instances provided by
``gce''. However it is worth mentioning that the ``gce'' ``large''
\emph{VM\_size} performs similarly to the ``aws'' ``large'',''cpuopt''
and ``ioopt'' \emph{VM\_size}. Overall ``gce'' provides more predictable
performance than the ``aws'' cloud. Finally, as in the ``aws''
case, we could not find any evidence that ``day-of-the-week'' and
``time\_of\_the\_day'' affects the QoS for ``gce'' and ``aws''
clouds. 
\begin{table}
\caption{QoS value states representation using hierarchal discretization for
$\theta_{compile}$. }

\centering{}%
\begin{tabular}{|c|c|c|}
\hline 
State & Range & Counts\tabularnewline
\hline 
\hline 
1 & 0 to 41 & 124\tabularnewline
\hline 
2 & 41 to 233 & 4910\tabularnewline
\hline 
3 & 213 to 405 & 1230\tabularnewline
\hline 
4 & 405 to 701 & 1007\tabularnewline
\hline 
5 & 701 to 784 & 19\tabularnewline
\hline 
6 & 784 to 918 & 7\tabularnewline
\hline 
7 & 918 to 1046 & 1\tabularnewline
\hline 
8 & 1046 to 1194 & 1\tabularnewline
\hline 
9 & 1194 to 1424 & 4\tabularnewline
\hline 
10 & 1424 to 1529 & 1\tabularnewline
\hline 
11 & 1529 to 1620 & 3\tabularnewline
\hline 
12 & 1620 to 2028 & 9\tabularnewline
\hline 
13 & 2028 to 2512 & 2\tabularnewline
\hline 
15 & 2654.5 and up & 1\tabularnewline
\hline 
\end{tabular}
\end{table}

\subsection{Memory Performance Diagnosis}

Hardware dominated applications not only depends on CPU but also on
memory. The memory dataset $(\theta_{memory})$ contains values related
to ``aws'' cloud and has 4581 rows in total. We again used hierarchical
discretisation method with manual fine tuning to discretize the QoS
values. In all, we created thirteen states for this dataset as shown
in Table IV. The aim for memory diagnosis is to determine the effect
of various factors on the memory dominated applications. Therefore,
in this case, we varied the states of all factors mentioned in Table
I. We started by selecting the ``micro'' \emph{VM\_size} in ``us''
\emph{region}. We found the performance of ``micro'' \emph{instance}
to be reasonably predictable where there was 78\% chance that the
values will lie in the range of 3612 and 3872 MB/sec (state 8). We
then varied the \emph{region} and selected ``eu'' and found an increase
in the performance not only in terns of bandwidth but also regarding
certainty. In particular, in this case, we found that most of the
QoS values lie in the range between 4116 and 4539 MB/sec (state 10)
with the probability of 87\%. We also found out that in this case,
``aws'' mostly employed the ``Intel Xeon E5\_2650 2GHz'' CPU with
the probability of more than 80\% in both ``us'' and ``eu'' \emph{regions}.
\begin{table}
\caption{QoS value states representation using hierarchal discretization for
$\theta_{memory}$.}

\centering{}%
\begin{tabular}{|c|c|c|}
\hline 
State & Range & Counts\tabularnewline
\hline 
\hline 
1 & 1 to 1039 & 135\tabularnewline
\hline 
2 & 1039 to 1425 & 61\tabularnewline
\hline 
3 & 1425 to 1909 & 549\tabularnewline
\hline 
4 & 1909 to 2318 & 569\tabularnewline
\hline 
5 & 2318 to 2577 & 1\tabularnewline
\hline 
6 & 2577 to 3205 & 20\tabularnewline
\hline 
7 & 3205 to 3612 & 35\tabularnewline
\hline 
8 & 3612 to 3872 & 490\tabularnewline
\hline 
9 & 3872 to 4116 & 127\tabularnewline
\hline 
10 & 4116 to 4539 & 551\tabularnewline
\hline 
11 & 4539 to 5101  & 84\tabularnewline
\hline 
12 & 5101 to 5651 & 969\tabularnewline
\hline 
13 & \emph{greater than 5651} & 990\tabularnewline
\hline 
\end{tabular}
\end{table}

We then studied the performance of \textquotedblleft small\textquotedblright{}
\emph{VM\_size} and its effects on the QoS value. As in the previous
cases, this instance provided lower performance compared to the \textquotedblleft micro\textquotedblright{}
instance in both the \emph{regions}. In the case of the \textquotedblleft eu\textquotedblright{}
\emph{region}, most of QoS values (93\% probability) lie in the range
of 1909 and 2318 MB/sec (state 4). In the \textquotedblleft us\textquotedblright{}
\emph{region}, nearly 79\% of the QoS values lie in the range of 1425
to 1909 MB/sec (state 3). The rest lie in lower ranges, i.e., between
1 and 1425 MB/sec (states 1 and 2). The lower performance of \textquotedblleft aws\textquotedblright{}
\emph{VM\_size} in both the \emph{regions} is attributed to the fact
that \textquotedblleft aws\textquotedblright{} consistently deploys
VMs on one of the better-performing CPUs in \textquotedblleft eu\textquotedblright ;
whereas, in the \textquotedblleft us\textquotedblright{} \emph{region},
other \emph{CPU} types are also considered with a higher probability. 

We also studied the performance of the \textquotedblleft large\textquotedblright{}
VM\_size and their effects on QoS value. We found out that even in
this case (as with CPU and OLTP), these instance provides better and
more predictable performance. For instance, \textquotedblleft large\textquotedblright{}
VM\_size in the \textquotedblleft us\textquotedblright{} region can
support QoS values in the range of 5101 to 5651 MB/sec (state 12)
with 93\% probability. Further the same instance, in the \textquotedblleft eu\textquotedblright{}
region supports even higher QoS values that lie in the range of 5651
and 6316.1 MB/sec. It is worth noting that \textquotedblleft aws\textquotedblright{}
employs the same CPU (\textquotedblleft Intel E5\_2670 2.50 GHz) in
both the regions for \textquotedblleft large\textquotedblright{} instances,
leading to higher performance. 

The $\theta_{memory}$ dataset also contains values for ``ioopt''
and ``cpuopt'' specialised instances for the ``eu'' \emph{region}.
We diagnosed the performance for both the instances and found that
none of these instances match the performance of the ``large'' \emph{VM\_size}.
For example, for the ``ioopt'' case, there is greater than 74\%
chance that the QoS values will lie above 5101 MB/sec (state 11),
and there is 21\% chance that the QoS values will lie in the range
of 3872 and 4116 MB/sec (state 9). Similarly, for the ``cpuopt''
case, there is appox. 81\% probability that the QoS values will lie
above 5101 MB/sec (state 12), where there is approx. 79\% chance that
these values will lie above 5651 MB/sec (states 12); the rest of the
QoS values mainly lie in the range of 4539 and 5101 MB/sec. Finally,
as in the previous cases, we did not find any evidence that ``day-of-the-week''
and ``time-of-the-day'' has any impact on any other parameter in
a BN.

\subsection{OLTP Performance Diagnosis}

The OLTP benchmark aims to study the performance related to multi-tenancy
in cloud systems. From Table 1, we note that in this dataset, there
are 3969 entries for this dataset $(\theta_{OLTP})$. The low number
of values corresponds to the data regarding to EC2 cloud. This data
set does not contain values related to the GCE. As can be observed
from Table 1, for this benchmark, the QoS values are widely distributed
with 95\% of the data lying in the range of 0 queries/sec to 1000
queries/sec, and with the standard deviation of 281.74 queries/sec.
This variation in the QoS values can be attributed to the fact that
multi-tenancy leads to low performance and leads to unpredictable
behaviour \cite{Leitner2016}. As in the CPU diagnosis case mentioned
above, for OLTP diagnosis, we created and tested several BNs. Our
aim was to study the effect of several factors on each other and most
importantly, on the OLTP QoS values. As QoS values were continuous,
we discretized them into finite states of different sizes. We used
hierarchical discretization method and discretized the OLTP QoS values
into three states with different counts as shown in Table V. As can
be observed from the Table, most of the QoS values lie in the range
of 0 to 196 queries/sec. This followed by the range of 196 to 561
queries/sec, and lastly, the range of 561 to 1130 where only 33 values
exist. 

To study the impact of several factors on the QoS Value, we first
selected the \textquotedblleft us\textquotedblright{} \emph{region},
\textquotedblleft aws\textquotedblright{} as the cloud, and varied
the \emph{VM\_size} as \textquotedblleft micro\textquotedblright ,
``small\textquotedblright , ``large\textquotedblright . As discussed
previously, these selections were entered as evidence $(e\in\,E)$
in a BN. We studied several hypotheses such as \emph{\textquotedblleft large
VMs provide better QoS values\textquotedblright }. In this case, the
larger VM should increase the throughput in queries/sec. Firstly,
we tested this hypothesis with \textquotedblleft micro\textquotedblright{}
\emph{VM\_size} and \textquotedblleft us\textquotedblright{} \emph{region}
to determine the QoS value and \emph{CPU}. After performing the inference,
we found out that nearly 98\% of the QoS values lie in state 1, i.e.,
between the range of 0 to 196 queries/sec. We also inferred that the
\textquotedblleft micro\textquotedblright{} \emph{VM\_size} in the
\textquotedblleft aws\textquotedblright{} \textquotedblleft us\textquotedblright{}
cloud mainly (82\% probability) uses the \textquotedblleft Intel Xeon
2650 cpu with 2 GHz\textquotedblright{} \emph{CPU}. We then tested
the same hypothesis by only changing the evidence as \textquotedblleft small\textquotedblright{}
for the factor \emph{VM\_size}. We noticed no change in the QoS value
compared to the ``micro'' \emph{VM\_size}, leading us to believe
that in the case of OLTP benchmark, ``micro'' and ``small'' \emph{VM\_size}
perform rather similarly; with 78\% probability Intel Xeon 2650 \emph{CPU}
with 2 GHz processor was used for the ``small'' \emph{VM\_size}
as well. In this case, our diagnosis is not absolute, rather based
on the limited dataset and the variability of data, we reached this
conclusion. We assert that this OLTP based benchmarking should be
done for a longer duration to build a larger dataset to retest this
hypothesis.

We again tested the same hypothesis but now by keeping all the evidences
fixed and by only varying the state of the factor \emph{VM\_size}
to \textquotedblleft large\textquotedblright . From this test, we
inferred that QoS value increases and lies mostly in the range of
196 to 561 queries/sec (state 2) validating the hypothesis that larger
\emph{VM\_size} provide better QoS performance. The \emph{VM\_size}
also contains two other states namely \textquotedblleft cpuopt\textquotedblright{}
and \textquotedblleft ioopt\textquotedblright{} representing CPU and
IO optimised VMs in the dataset. To verify whether I/O optimised \emph{VM\_size}
leads to further QoS performance improvement, we kept all the evidences
fixed but varied the state of the \emph{VM\_size} to \textquotedblleft ioopt\textquotedblright .
After inference, we concluded that \textquotedblleft ioopt\textquotedblright{}
instance provided the best QoS values with most of values (with 93\%
probability) lying in the range of 561 queries to 1130 queries/sec
(state 3). We also found out that the \textquotedblleft ioopt\textquotedblright{}
\emph{VM\_size} employs a more powerful \textquotedblleft Intel Xeon
E5\_2670 2.50 Ghz\textquotedblright{} CPU. 

To study the impact of \emph{region} on the OLTP QoS values, we studied
the same hypothesis by changing the state of \emph{region} from \textquotedblleft us\textquotedblright{}
to \textquotedblleft eu\textquotedblright . We then performed inference
one by one by selecting the state of\emph{ VM\_size} from \textquotedblleft micro\textquotedblright ,
to \textquotedblleft ioopt\textquotedblright , our analyses led us
to conclude that OLTP performance remain rather stable across both
\emph{regions} for \textquotedblleft micro\textquotedblright , \textquotedblleft small\textquotedblright ,
and \textquotedblleft large\textquotedblright{} \emph{VM\_size}. We
found that this dataset do not contain values related to \textquotedblleft ioopt\textquotedblright{}
\emph{VM\_size} for \textquotedblleft us\textquotedblright{} \emph{regions}.
Interestingly, we also concluded that in the \textquotedblleft eu\textquotedblright{}
\emph{region}, more expensive \textquotedblleft cpuopt\textquotedblright{}
\emph{VM\_size} performs similarly to \textquotedblleft large\textquotedblright{}
\emph{VM\_size}. Lastly, through Bayesian diagnosis, we inferred that
\textquotedblleft time-of-the-day\textquotedblright{} and \textquotedblleft day-of-the-week\textquotedblright{}
do not affect any other RV significantly. 
\begin{table}
\caption{QoS value states representation using hierarchal discretization for
$\theta_{oltp}$. }

\centering{}%
\begin{tabular}{|c|c|c|}
\hline 
State & Range & Counts\tabularnewline
\hline 
\hline 
1 & 0 to 196 & 2152\tabularnewline
\hline 
2 & 196 to 561 & 1327\tabularnewline
\hline 
3 & 561 to 1130 & 33\tabularnewline
\hline 
\end{tabular}
\end{table}

\subsection{I/O Performance Diagnosis}

The I/O benchmark also aims to study the performance related to multi-tenancy
in cloud systems. From Table 1, we note that there were 7377 data
points present in the dataset ($\theta_{IO}$) representing the values
for ``aws'' and ``gce'' clouds. We did not find any significant
variation in the QoS values. As in the previous cases, we discretized
the I/O QoS values which were continuous, into finite states of different
sizes (see Table 5) We first analysed the performance of ``aws''
cloud by varying parameters listed above. Initially, we selected the
``micro'' \emph{VM\_size}' in the ``us'' \emph{region} and found
that most of the QoS values (77\% chance) lie in the range of 0 and
2 Mb/sec (state 1). We then varied the \emph{region} to ``eu'' and
found similar results albeit with less predictability, where there
is with only 66\% chance that the values will lie in this range. We
then varied the \emph{VM\_size} to ``small'' and found nearly no
change in the result. Rather the QoS values become less predictable
in the ``us'' \emph{region} with close to half of the values lie
in with states 1 and 2. In the ``eu'' \emph{region}, the value were
widely distributed with 53\% chance that QoS will lie in state 2,
followed by 28\% chance in state 1 and 18\% chance that they will
lie in state 3, respectively. 

Again, in this case, we found that the ``gce'' \emph{cloud} provides
significantly high predictable values compared to the ``aws'' \emph{cloud}.
In that, we concluded that ``gce'' and ``micro'' \emph{VM\_size}
will lead to state 1 with 99.5\% chance in both ``us'' and ``eu''
\emph{regions}. Similarly, in the case of ``small'' \emph{VM\_size}
in the ``eu'' \emph{region}, there is 100\% chance that the QoS
values will lie in state 2. The performance for ``gce'' \emph{cloud}
in the ``us'' \emph{region} was less predictable with only 71\%
chance that the QoS values will lie in state 2 and rest in state 1,
respectively. In the case of the ``large'' \emph{VM\_size}, ``aws''
\emph{cloud} provided more predictable results in this case where
there was an average 80.5\% chance that the QoS values will lie in
state 3, and the rest of the values will lie in state 2. In the case
of ``gce'', there was only 67\% chance that the QoS values will
lie in state 3 and rest of the values will lie in state 2. 

This dataset also contains QoS values for ``oopt'' and ``cpuopt''
\emph{VM\_size} for ``aws'' \emph{cloud}. The ``oopt'' \emph{VM\_size}
performs very well with 100\% chance that the values will lie in state
3. The ``cpuopt'' \emph{VM\_size} performed rather poorly with only
55\% chance that the QoS values will lie in state 3 and rest of the
values will lie in state 2. Again even in this case, we did not find
any conclusive evidence that ``time-of-the-day'' and ``day-of-the-week''
factors have any significant impact on the QoS values for all the
clouds. 
\begin{table}
\caption{QoS value states representation using hierarchal discretization for
$\theta_{IO}$. }

\centering{}%
\begin{tabular}{|c|c|c|}
\hline 
State & Range & Counts\tabularnewline
\hline 
\hline 
1 & 0 to 2 & 2461\tabularnewline
\hline 
2 & 2 to 17 & 2457\tabularnewline
\hline 
3 & 17 to 1009.6 & 2459\tabularnewline
\hline 
\end{tabular}
\end{table}

\subsection{Cloud QoS Prediction}

The previous section validated ALPINE's cloud performance diagnosis
capability under uncertainty. This section presents the results related
to cloud QoS prediction. As referred to in section 2, a BN can be
modelled in many ways. It can be a simple Naive Bayes Model (NBN)
(see Fig 2(a) where all the factors are conditionally independent
given an outcome, i.e., QoS value. Alternatively, it can be a more
complex BN (CBN) (See Fig. 2 (d)) where more arcs between the factors
are connected to determine more complex relationships between them.
Fig. 2 (c) shows another simple model; this is a Noisy-Or model (NOR)
where all the factors directly affect the QoS value. Finally, Fig.
2 (b) presents a Tree-augmented Naive Bayes Model (TAN); this model
is similar to NBN. However, in this model, more arcs are connected
to determine more complex relationships between the factors. All of
these models were learned after we performed discretization on the
raw QoS values. To validate BNs prediction accuracy, we used 10-fold
cross-validation which is a widely accepted method to determine the
accuracy and correctness of a model \cite{russelandnorvig,Caqoem15}.
For training the model, we again used the EM algorithm \cite{Leitner2016}.
Table VII shows the prediction accuracy of all BNs. We conclude that
BNs can predict QoS efficiently with an overall prediction accuracy
of approximately 91.93\%, which is an excellent result. To our surprise,
we found that even the simplest BNs could achieve high prediction
accuracy (compared to CBN) using the dataset \cite{LeitnerC14,Leitner2016}
utilised in this paper. The low prediction accuracy in the case of
I/O dataset $(\theta_{IO})$ was because of a very narrow distribution
of I/O QoS values. We assert that these results can be beneficial
for the stakeholders for not only the best cloud selection but also
to predict the QoS that their application might perceive by using
a combination of factors mentioned above. 
\begin{table}
\caption{Cloud QoS Prediction accuracy (\%) for different type of Bayesian
Networks.}

\centering{}%
\begin{tabular}{|c|c|c|c|c|c|}
\hline 
BN Type & CPU & Compile & Memory & OLTP & I/O\tabularnewline
\hline 
\hline 
NBN & 97.12 & 95.93 & 89.54 & 97.40 & 76.21\tabularnewline
\hline 
TAN & 99.24 & 96.08 & 92.20 & 97.40 & 76.17\tabularnewline
\hline 
NOR & 99.24 & 95.65 & 91.42 & 97.40 & 76.08\tabularnewline
\hline 
CBN & 99.24 & 96.09 & 92.70 & 97.40 & 76.04\tabularnewline
\hline 
\end{tabular}
\end{table}

\section{Conclusion and Future Work}

This paper proposed, developed and validated ALPINE - a Bayesian system
for cloud performance diagnosis and prediction. The results presented
in the paper clearly demonstrate that ALPINE can be used for efficiently
diagnose cloud performance even in the case of limited data. The major
highlight of ALPINE is that it can consider several factors simultaneously
(CPU, VM size, regions, cloud providers, type of benchmark, time-of-the-day,
day-of-the-week, and QoS values) for the root-cause diagnosis of cloud
performance. In particular, a stakeholder can enter the evidence regarding
multiple factors to determine their impact on other factors. The state-of-the-art
methods lack this capability. ALPINE can model complex and uncertain
relationships between these factors probabilistically to reason about
several hypotheses regarding cloud performance. We also validated
ALIPNE's prediction performance and showed that it achieves an overall
prediction accuracy of 91.93\%. Therefore, we assert that stakeholders
can use ALPINE for efficient cloud ranking, selection, and orchestration.
As a future work, we will collect more data for several other cloud
providers.\bibliographystyle{plain}
\bibliography{IEEEfull,/xUsers/karan/Documents/JMD_RESEARCH/bibs/JMD_Mobile_Cloud_Computing}

\end{document}